\documentclass[preprint,12pt]{elsarticle}

\usepackage{amssymb,amsmath}
\usepackage{url}

\newcommand{\ket}[1]{| \, #1 \, \rangle}

\journal{Nuclear Physics A}

\begin{document}

\begin{frontmatter}



\title{Antikaon-nucleon interaction and $\Lambda(1405)$ \\
in chiral SU(3) dynamics}


\author[a]{Yuki~Kamiya}
\author[b]{Kenta~Miyahara}
\author[c]{Shota~Ohnishi}
\author[d]{Yoichi~Ikeda} 
\author[a]{Tetsuo~Hyodo\corref{cor1}}
\ead{hyodo@yukawa.kyoto-u.ac.jp}
\author[e]{Eulogio~Oset}
\author[f]{Wolfram~Weise}
\address[a]{Yukawa Institute for Theoretical Physics, Kyoto University, Kyoto 606-8502, Japan}
\address[b]{Department of Physics, Graduate School of Science, Kyoto University, Kyoto 606-8502, Japan}
\address[c]{Department of Physics, Hokkaido University, Sapporo 060-0810, Japan}
\address[d]{RIKEN Nishina Center, Wako, Saitama 351-0198, Japan}
\address[e]{Departamento de F\'{\i}sica Te\'orica and IFIC, Centro Mixto Universidad de Valencia-CSIC,
Institutos de Investigaci\'on de Paterna, Aptdo. 22085, 46071 Valencia,
Spain}
\address[f]{Physik Department, Technische Universit\"{a}t M\"{u}nchen, D-85747 Garching, Germany}
\cortext[cor1]{Corresponding author}

\begin{abstract}
The properties of the $\Lambda(1405)$ resonance are key ingredients for determining the antikaon-nucleon interaction in strangeness nuclear physics, and the novel internal structure of the $\Lambda(1405)$ is of great interest in hadron physics, as a prototype case of a baryon that does not fit into the simple three-quark picture. We show that a quantitative description of the antikaon-nucleon interaction with the $\Lambda(1405)$ is achieved in the framework of chiral SU(3) dynamics, with the help of recent experimental progress. Further constraints on the $\bar{K}N$ subthreshold interaction are provided by analyzing $\pi\Sigma$ spectra in various processes, such as the $K^{-}d\to \pi\Sigma n$ reaction and the $\Lambda_{c}\to \pi\pi\Sigma$ decay. The structure of the $\Lambda(1405)$ is found to be dominated by an antikaon-nucleon molecular configuration, based on its wavefunction derived from a realistic $\bar{K}N$ potential and the compositeness criteria from a model-independent weak-binding relation.

\end{abstract}

\begin{keyword}


$\bar{K}N$ interaction \sep $\Lambda(1405)$ resonance \sep Chiral SU(3) dynamics\sep $\pi\Sigma$ spectrum \sep Compositeness

\end{keyword}

\end{frontmatter}



\section{Introduction}

Understanding the $\Lambda(1405)$ resonance has been a central issue in hadron-nuclear physics in the strangeness sector. There have been plenty of hadron spectroscopy investigations to clarify the internal structure of the $\Lambda(1405)$ using various pictures: three-quark excited state~\cite{Isgur:1978xj}, $\bar{K}N$ molecular structure~\cite{Dalitz:1959dn,Dalitz:1960du,Dalitz:1967fp}, and so on. Among others, the $\bar{K}N$ molecule picture of the $\Lambda(1405)$ is of particular interest. In this picture, the $\Lambda(1405)$ is regarded as a quasi-bound state generated by the attractive $\bar{K}N$ interaction, which decays via the transition to the $\pi\Sigma$ channel. 

The key properties (mass and width) of the $\Lambda(1405)$ are the basic clues to identify its internal structure. In addition, because the $\Lambda(1405)$ is located below the $\bar{K}N$ threshold, a precise determination of the $\Lambda(1405)$ properties sets constraints on the behavior of the $\bar{K}N$ interaction in the subthreshold region, which is relevant for the study of possible quasi-bound antikaonic nuclei ~\cite{PL7.288,Akaishi:2002bg}. For future orientation of strangeness nuclear physics, we summarize here the current status of the $\Lambda(1405)$ and the $\bar{K}N$ interaction. At the same time, the $\bar{K}N$ molecular picture itself requests a critical examination. We thus raise the following questions:
\begin{enumerate}
\item To what precision are the properties of the $\Lambda(1405)$ determined?
\item What will be the next steps to further constrain the $\bar{K}N$ interaction?
\item How can the $\bar{K}N$ molecular picture of the $\Lambda(1405)$ be verified?
\end{enumerate}
In this paper, we would like to address these questions in the framework of chiral SU(3) dynamics~\cite{Kaiser:1995eg,Oset:1998it,Oller:2000fj,Lutz:2001yb,Hyodo:2011ur}, combined with recent precise experimental measurements.

\section{Chiral SU(3) dynamics for $\bar{K}N$ scattering}

Chiral symmetry is the guiding principle to study the low-energy phenomena of nonperturbative QCD. As a consequence of the spontaneous breaking of chiral symmetry, low-energy theorems determine the interaction and dynamics of the corresponding Nambu-Goldstone bosons~\cite{Weinberg:1966kf,Tomozawa:1966jm}. Chiral perturbation theory~\cite{Weinberg:1979kz,Gasser:1983yg} provides a way to elaborate the results of the low-energy theorems based on a systematic power counting scheme. This theory is quite successful in describing the dynamics of pions in the two-flavor sector~\cite{Bernard:1995dp,Scherer:2012xha}.

In the three-flavor sector, there appear kaons and the eta meson associated with the spontaneous breaking of chiral SU(3)$_{R}\times$SU(3)$_{L}$ symmetry. In this case, the relatively large strange quark mass causes an interplay between spontaneous and explicit chiral symmetry breaking. In fact, the existence of the $\Lambda(1405)$ resonance near the threshold indicates the strong nonperturbative dynamics of the $\bar{K}N$ system. Chiral SU(3) dynamics~\cite{Kaiser:1995eg,Oset:1998it,Oller:2000fj,Lutz:2001yb,Hyodo:2011ur} offers a non-perturbative coupled-channels framework in which the constraints from chiral symmetry are properly encoded.

\subsection{$\Lambda(1405)$ and $\bar{K}N$ scattering}

We start from the chiral SU(3)$_{R}\times$SU(3)$_{L}$ meson-baryon effective Lagrangian 
\begin{align} 
   {\cal L}_{\rm eff}({\cal B, U}) 
   = {\cal L}_M({\cal U}) 
   + {\cal L}_{MB}^{(1)}({\cal B,U}) 
   +  {\cal L}_{MB}^{(2)}({\cal B,U}),
   \label{eq:Lagrangian}
\end{align}
where the matrix fields ${\cal U}$ and ${\cal B}$ represent the octet pseudoscalar mesons $(\pi, K, \bar{K}, \eta)$ and the octet baryons $(N, \Lambda, \Sigma, \Xi)$, respectively. The leading order ${\cal O}(p)$ meson-baryon vertices are incorporated in ${\cal L}_{MB}^{(1)}({\cal B,U})$, which includes the meson-baryon four-point couplings (Weinberg-Tomozawa term), and the three-point Yukawa couplings with low-energy constants $D$ and $F$. In the next-to-leading order (NLO) ${\cal O}(p^2)$ Lagrangian ${\cal L}_{MB}^{(2)}({\cal B,U})$, the terms relevant for meson-baryon scattering are given with seven low-energy constants [$b_0$, $b_D$, $b_F$, and $d_i$ ($i=1,...,4$)].

We consider $s$-wave $\bar{K}N$ scattering with the chiral Lagrangian~\eqref{eq:Lagrangian}. In chiral perturbation theory, the diagrams for the scattering amplitude are classified according to the counting rule in powers of momentum $p$. The perturbative meson-baryon scattering amplitude $V_{ij}(W)$ at the total energy $W$ up to ${\cal O}(p^{2})$ is given by
\begin{align} 
   V_{ij}(W) = V^{\rm WT}_{ij}(W)+V^{\rm s}_{ij}(W)
   +V^{\rm u}_{ij}(W)+V^{\rm NLO}_{ij}(W)
   \label{eq:interaction} ,
\end{align}
where $i,j$ denotes the channel indices, $\bar{K}N$, $\pi\Lambda$, $\pi\Sigma$, $\eta\Lambda$, $\eta\Sigma$, and $K\Xi$. There are two types of the leading order ${\cal O}(p)$ contributions: the Weinberg-Tomozawa contact term $V^{\rm WT}_{ij}(W)$ and the $s$- and $u$-channel Born terms $V^{\rm s}_{ij}(W)$ and $V^{\rm u}_{ij}(W)$ which consist of the Yukawa vertices and the baryon propagator. Because the Born terms mainly contribute to $p$-wave scattering, the Weinberg-Tomozawa term gives the dominant contribution for $s$-wave scattering. The vertices from the NLO Lagrangian ${\cal L}_{MB}^{(2)}({\cal B,U})$ provide higher order ${\cal O}(p^{2})$ contributions $V^{\rm NLO}_{ij}(W)$ at tree level. One-loop diagrams appear at the order ${\cal O}(p^{3})$. Explicit expressions can be found in Ref.~\cite{Hyodo:2011ur}.

The perturbative amplitude~\eqref{eq:interaction} cannot describe the $\Lambda(1405)$ below the $\bar{K}N$ threshold. This is a situation analogous to the chiral effective field theory approach for the nuclear force~\cite{Epelbaum:2008ga,Machleidt:2011zz} where the deuteron exists below the two-nucleon threshold. To overcome this difficulty, it is mandatory to build a nonperturbative $\bar{K}N$ scattering amplitude. This has been achieved by solving the Lippmann-Schwinger equation~\cite{Kaiser:1995eg,Oset:1998it}, or by constructing the general form of the amplitude in the N/D method~\cite{Oller:2000fj}. In both cases, the nonperturbative coupled-channel scattering amplitude $T_{ij}(W)$ satisfies the following matrix equation in coupled channels:
\begin{align} 
   T_{ij}(W) = V_{ij}(W)+V_{ik}(W)G_{k}(W)T_{kj}(W) .
   \label{eq:amplitude} 
\end{align}
The loop function $G_{k}(W)$, with ultraviolet divergences removed, contains subtraction constants for each channel. In the $\bar{K}N$ scattering sector six subtraction constants are introduced: $a_{\bar{K}N}$, $a_{\pi\Lambda}$, $a_{\pi\Sigma}$, $a_{\eta\Lambda}$, $a_{\eta\Sigma}$, and $a_{K\Xi}$.

The low-energy constants and the subtraction constants in the amplitude are determined by reproducing experimental data such as the total cross sections of $K^{-}p$ scattering into elastic and inelastic channels, and the threshold branching ratios. It turns out that the phenomenologically successful description of these data can be obtained mainly by the Weinberg-Tomozawa term $V^{\rm WT}(W)$~\cite{Oset:1998it}. At the same time, the signature of the $\Lambda(1405)$ appears in the $\pi\Sigma$ mass spectrum.

\begin{figure}[tb]
  \centerline{
  \includegraphics[width=9cm,bb=0 0 727 675]{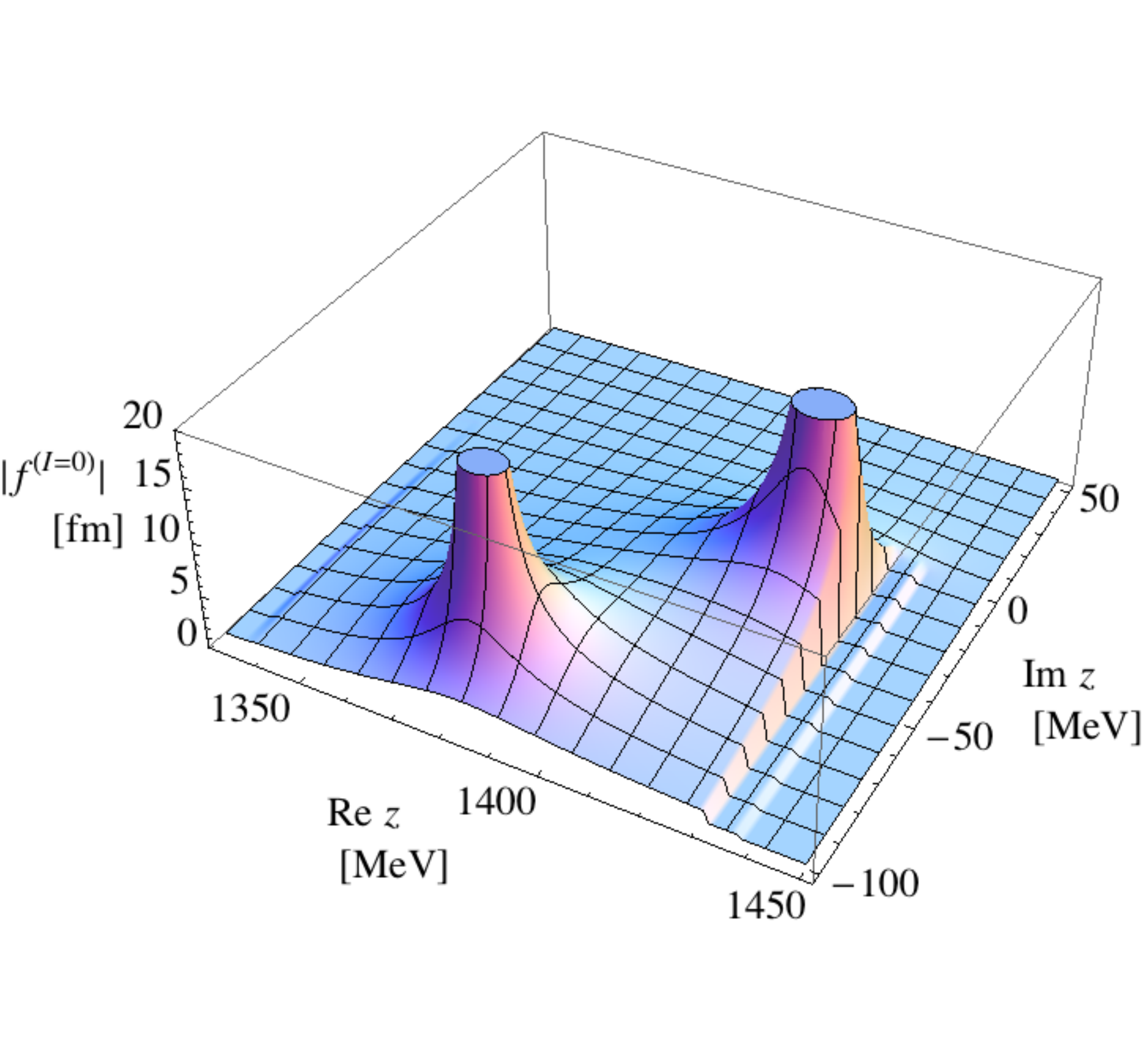} 
  }
  \caption{Absolute value of the $I=0$ combination of the $\bar{K}N$ 
  scattering amplitude $|f^{(I=0)}(z)|$ defined in 
  Eq.~\eqref{eq:amplitudeI0} from the NLO approach in 
  Refs.~\cite{Ikeda:2011pi,Ikeda:2012au} in the complex energy $z$ plane.
  We choose the Riemann sheet which is most adjacent to the real axis.}
  \label{fig:pole}
\end{figure}

An interesting outcome is the novel pole structure of the $\Lambda(1405)$~\cite{Oller:2000fj}. Through the analytic continuation of the scattering amplitude in the complex energy plane, $W\to z$, two poles are found below the $\bar{K}N$ threshold, as shown in Fig.~\ref{fig:pole}. One pole is located near the $\bar{K}N$ threshold with a narrow width, and a second one appears at lower energy with a larger imaginary part. Because a pole singularity of the scattering amplitude corresponds to an eigenstate of the Hamiltonian, the appearance of the two poles indicates the existence of two independent states with the same quantum numbers. Extrapolation to the flavor SU(3) symmetric limit shows that these poles originate from the attractive Weinberg-Tomozawa forces acting in the flavor singlet and octet channels~\cite{Jido:2003cb}. In the isospin basis, the relevant attractive interactions are found in the $\bar{K}N$ and $\pi\Sigma$ channels~\cite{Hyodo:2007jq}, so the pole structure can be interpreted as the superposition of a $\bar{K}N$ quasi-bound state and a broad $\pi\Sigma$ resonance.

While the gross features of the $\Lambda(1405)$ in the $\bar{K}N$ amplitude are already well described just by the Weinberg-Tomozawa term, higher order contributions~\cite{Kaiser:1995eg,Lutz:2001yb} can be used to quantitatively improve the description. To determine the low-energy constants in the higher order terms, sufficiently many precise experimental data are necessary. In 2004, the DEAR collaboration reported measurements of kaonic hydrogen~\cite{Beer:2005qi} from which the $K^{-}p$ scattering length can be deduced (see the next section). Systematic studies with the NLO interactions however pointed out an inconsistency of the DEAR result with the scattering data~\cite{Borasoy:2004kk,Borasoy:2005ie,Oller:2005ig,Oller:2006jw}. 

\subsection{NLO analysis with precise kaonic hydrogen data}

An experimental breakthrough came in 2011 when the SIDDHARTA collaboration provided a new measurement of the shift $\Delta E$ and width $\Gamma$ of the 1s level of kaonic hydrogen~\cite{Bazzi:2011zj,Bazzi:2012eq}:
\begin{align} 
   \Delta E = 283\pm36(\mathrm{stat})\pm6(\mathrm{syst}) \textrm{ eV} , 
   \quad \Gamma = 541\pm89(\mathrm{stat})\pm22(\mathrm{syst})\textrm{ eV} .
   \label{eq:SIDDHARTA} 
\end{align}
Kaonic hydrogen is the Coulombic bound state of the $K^{-}p$ system. The 1s energy shift and width are induced by the strong interaction. In the non-relativistic effective Lagrangian approach, this shift and width are related to the $K^{-}p$ scattering length $a_{K^{-}p}$ as~\cite{Meissner:2004jr} 
\begin{align} 
   \Delta E - i\Gamma/2 
   = -2\alpha^3\,\mu_r^2\,a_{K^-p}\left[1+2\alpha\,\mu_r\,(1-\ln\alpha)\,a_{K^-p}\right] ,
   \label{eq:shiftwidth}
\end{align}
where $\alpha$ is the fine-structure constant and $\mu_r = m_{K^{-}} M_p/(m_{K^{-}} + M_p)$ is the $K^-p$ reduced mass. Thus the kaonic hydrogen measurement~\eqref{eq:SIDDHARTA} provides a direct constraint on the $\bar{K}N$ scattering amplitude at threshold.

The first systematic NLO analysis including the SIDDHARTA constraint has been performed in Refs.~\cite{Ikeda:2011pi,Ikeda:2012au}. The data base used in this analysis consists of the $K^{-}p$ total cross sections, the threshold branching ratios, and the $K^{-}p$ scattering length deduced from the SIDDHARTA data. We obtain a best fit result in the full NLO approach with $\chi^{2}/$d.o.f. $=0.96$, showing that the new measurement of kaonic hydrogen is now consistent with the scattering data. The same analysis can be performed with only the Weinberg-Tomozawa term. A reasonable fit is found with $\chi^{2}/$d.o.f. $=1.12$, provided that some of the subtraction constants are allowed to take unnatural values to compensate for higher order contributions. While this justifies, at least qualitatively, the use of Weinberg-Tomozawa model for the study of $\Lambda(1405)$, the shift and width are now obtained as $\Delta E=373$ eV and $\Gamma=495$ eV, outside the error bars of the energy shift in Eq.~\eqref{eq:SIDDHARTA}. The SIDDHARTA data are sufficiently precise to shed light on the small insufficiency of the leading-order (Weinberg-Tomozawa) approach.

\begin{figure}[tb]
  \centerline{
  \includegraphics[width=7cm,bb=0 0 360 252]{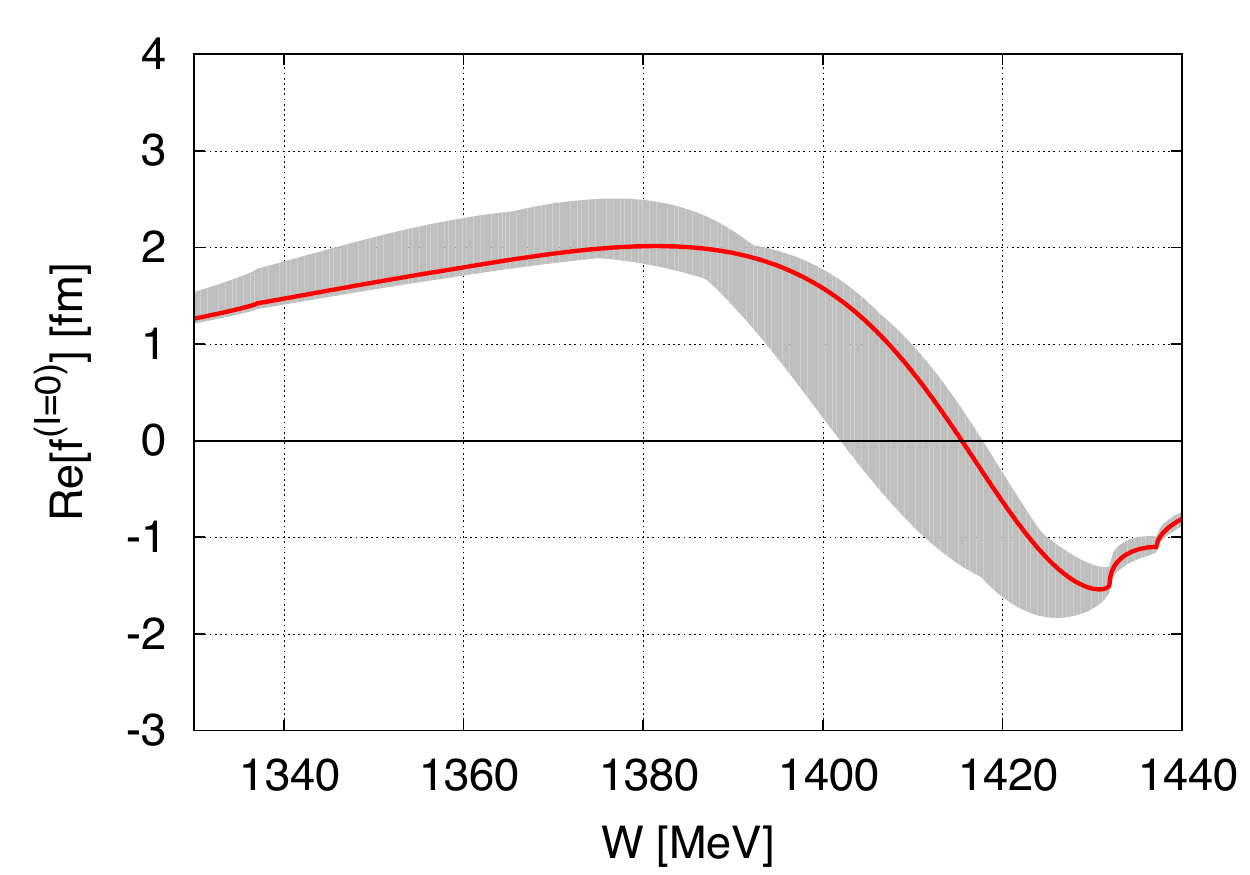}
  \includegraphics[width=7cm,bb=0 0 360 252]{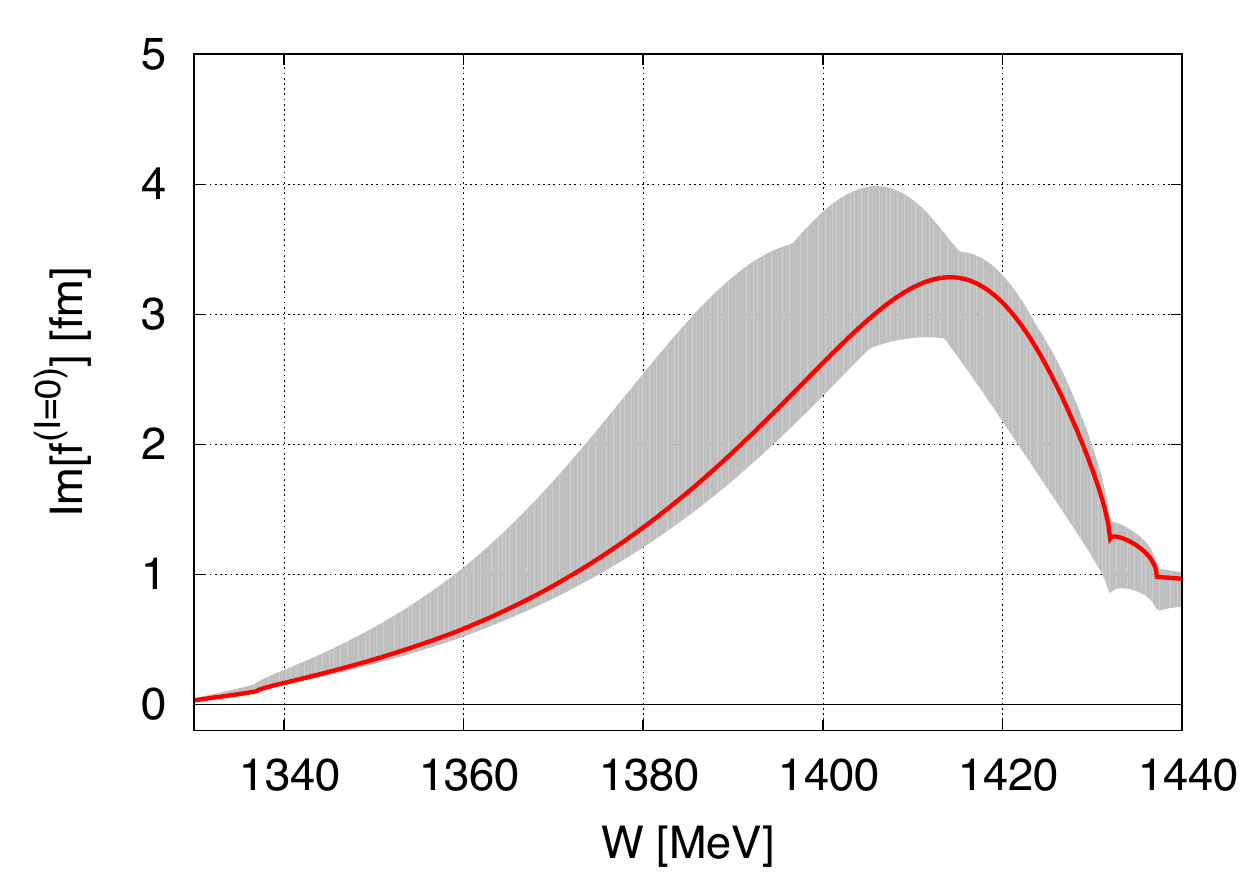}
  }
  \caption{Subthreshold extrapolation of the $I=0$ combination of the $\bar{K}N$ amplitude~\eqref{eq:amplitudeI0} in the NLO approach in Refs.~\cite{Ikeda:2011pi,Ikeda:2012au}. Shaded areas represent the uncertainty bands.}
  \label{fig:ampligudeI0}
\end{figure}

With a reliable $\bar{K}N$ scattering amplitude at hand, we are ready to perform an extrapolation into the subthreshold energy region. In Fig.~\ref{fig:ampligudeI0}, we show the subthreshold extrapolation of the $I=0$ combination of the $\bar{K}N$ amplitude in the NLO approach,
\begin{align} 
    f^{(I=0)}(W) 
    =
    \frac{1}{2}[
    f_{K^{-}p K^{-}p}(W)
    +2f_{K^{-}p \bar{K}^{0}n}(W)
    +f_{\bar{K}^{0}n\bar{K}^{0}n}(W)
    ] ,
    \label{eq:amplitudeI0}
\end{align}
with $f_{ij}(W)=-\sqrt{M_{i}M_{j}}\,T_{ij}(W)/(4\pi W)$. The uncertainty bands are evaluated according to the SIDDHARTA data~\eqref{eq:SIDDHARTA}. It turns out that the large uncertainty of the subthreshold extrapolation in the previous studies is significantly reduced by the precise determination of the $K^{-}p$ scattering length. In the complex energy plane, we find two poles at
\begin{align} 
    z_{1} = (1424^{+7}_{-23}- i \,26^{+3}_{-14})\, {\rm MeV},
    \quad 
    z_{2} = (1381^{+18}_{-6}- i \,81^{+19}_{-8})\, {\rm MeV} .
\end{align}
In this way, the existence of two poles is confirmed in the NLO analysis constrained by the SIDDHARTA data.

\begin{figure}[tb]
  \centerline{
  \includegraphics[width=7cm,bb=0 0 360 252]{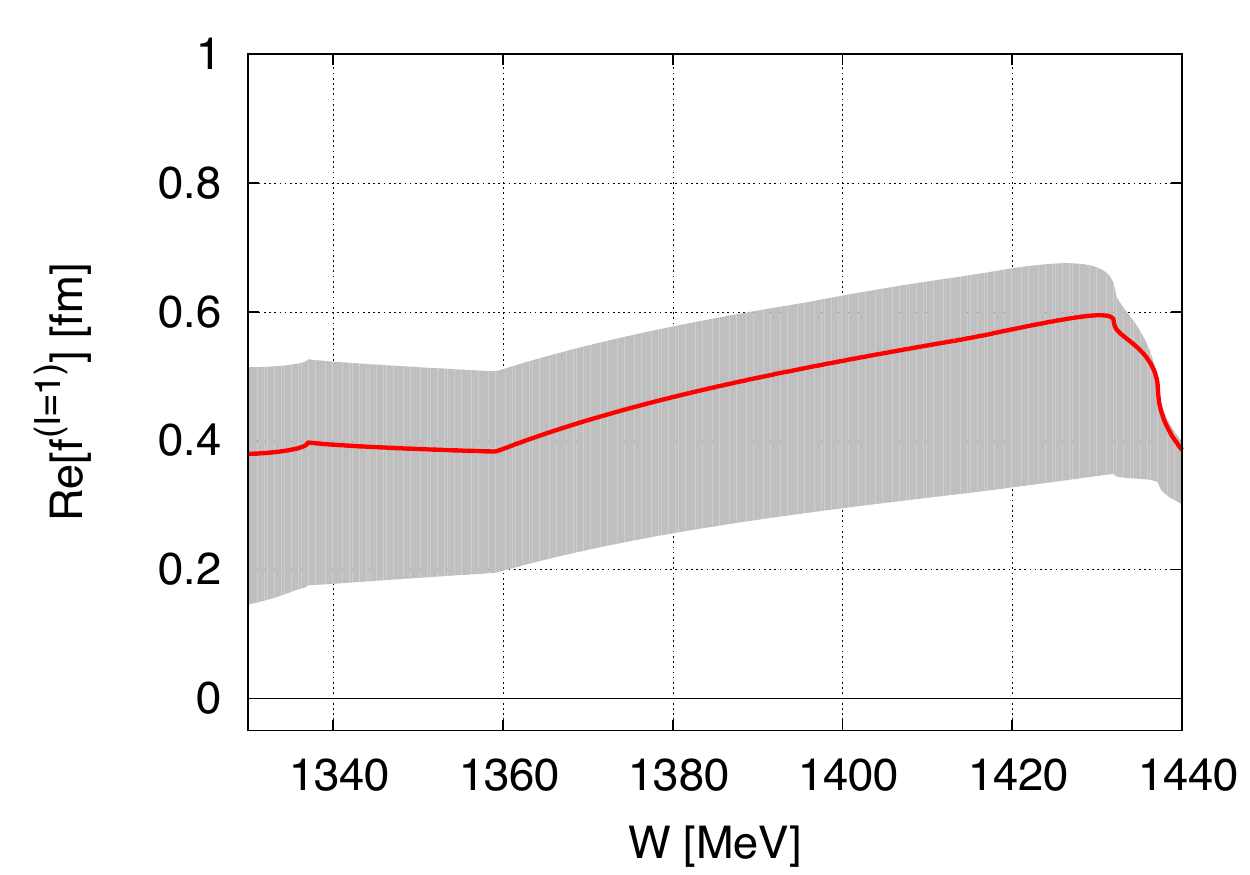}
  \includegraphics[width=7cm,bb=0 0 360 252]{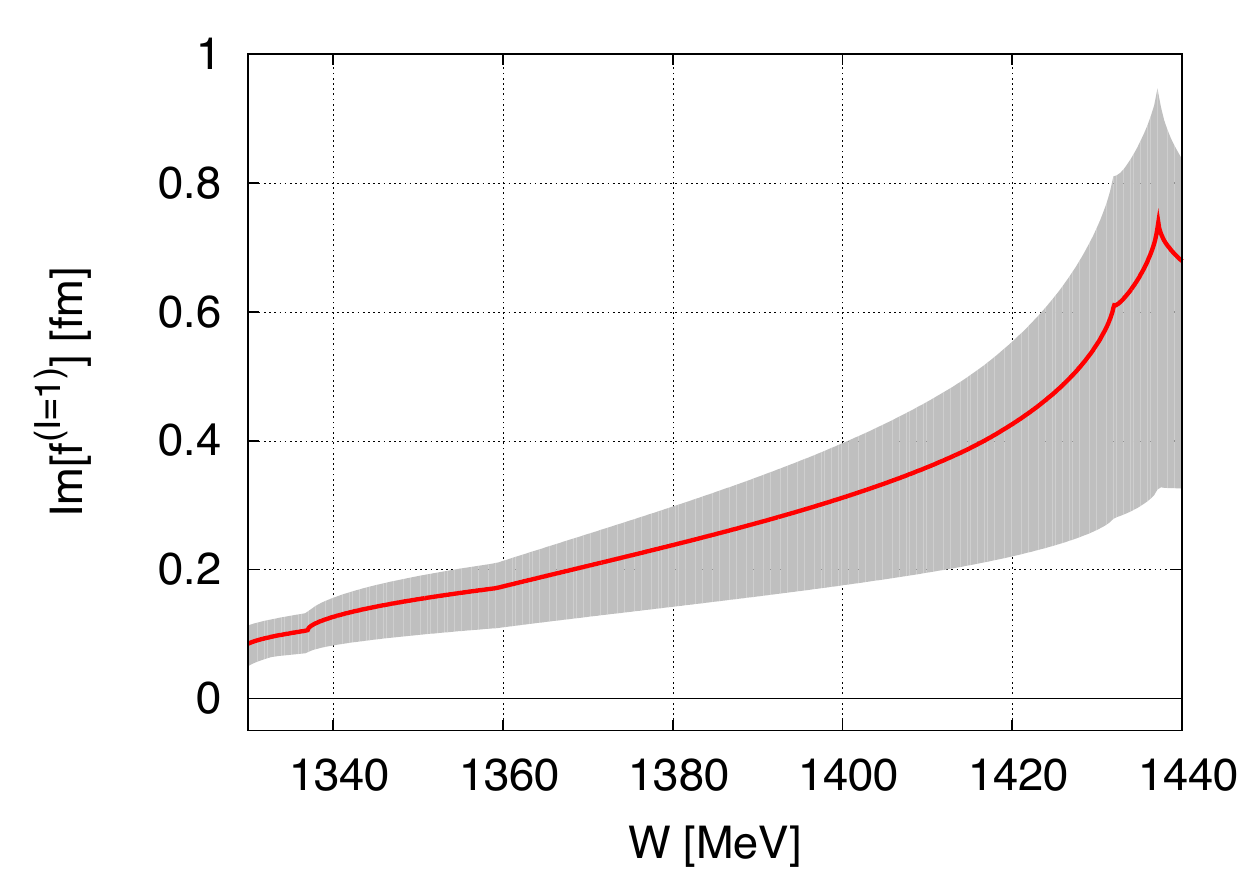}
  }
  \caption{Same as Fig~\ref{eq:amplitudeI1}, but for $I=1$.}
  \label{fig:ampligudeI1}
\end{figure}

So far we have focused on the $I=0$ amplitude in which the $\Lambda(1405)$ appears. In order to study the interaction of $\bar{K}$ in nuclei, we need both the $I=0$ and $I=1$ components. In Fig.~\ref{fig:ampligudeI1}, we plot the $I=1$ combination of the $\bar{K}N$ amplitude
\begin{align} 
    f^{(I=1)}(W) 
    =
    \frac{1}{2}[
    f_{K^{-}p K^{-}p}(W)
    -2f_{K^{-}p \bar{K}^{0}n}(W)
    +f_{\bar{K}^{0}n\bar{K}^{0}n}(W)
    ] .
    \label{eq:amplitudeI1}
\end{align}
Compared with the $I=0$ component, we observe a relatively larger uncertainty with respect to the central values. While the magnitude of the $I=1$ amplitude is smaller than that of the resonant $I=0$ amplitude, the uncertainty in the $I=1$ component should be reduced in the future. An important constraint on the $I=1$ amplitude is expected to come from measurements of the 1s energy shift and width of kaonic deuterium which involves both $K^{-}p$ and $K^{-}n$ interactions. It is interesting to note the cusp structure around the $\bar{K}N$ threshold of the $I=1$ amplitude. For practical purposes, this could be identified with a new state of $I=1$, similar to the $a_{0}(980)$ which shows up as a strong cusp around the $\bar{K}K$ threshold~\cite{Rubin:2004cq}. Based on phenomenology, a state of this type, and close in mass, has been claimed in Refs.~\cite{Gao:2010hy,Wu:2009nw}

\subsection{Pole structure in the $\Lambda(1405)$ energy region}\label{sec:polestructure}

To examine the systematic uncertainties of chiral SU(3) dynamics, we compare the results of Refs.~\cite{Ikeda:2011pi,Ikeda:2012au} in the previous section with those by other groups (see also Mini Review~\cite{Minireview} in Particle Data Group~\cite{Agashe:2014kda}). In Refs.~\cite{Guo:2012vv,Mai:2014xna}, the $\Lambda(1405)$ is studied in similar approaches with different treatments of the scattering equation and the fitting procedure. Here we summarize the studies that are based on the NLO chiral SU(3) interaction and perform the $\chi^{2}$ fitting analysis with uncertainty estimates including the SIDDHARTA data. Other recent studies of the $\Lambda(1405)$ can be found in Refs.~\cite{Cieply:2011nq,Roca:2013av,Roca:2013cca,Kamano:2014zba,Feijoo:2015yja,Kamano:2015hxa}.

\begin{figure}[tb]
  \centerline{
  \includegraphics[width=10cm,bb=0 0 360 216]{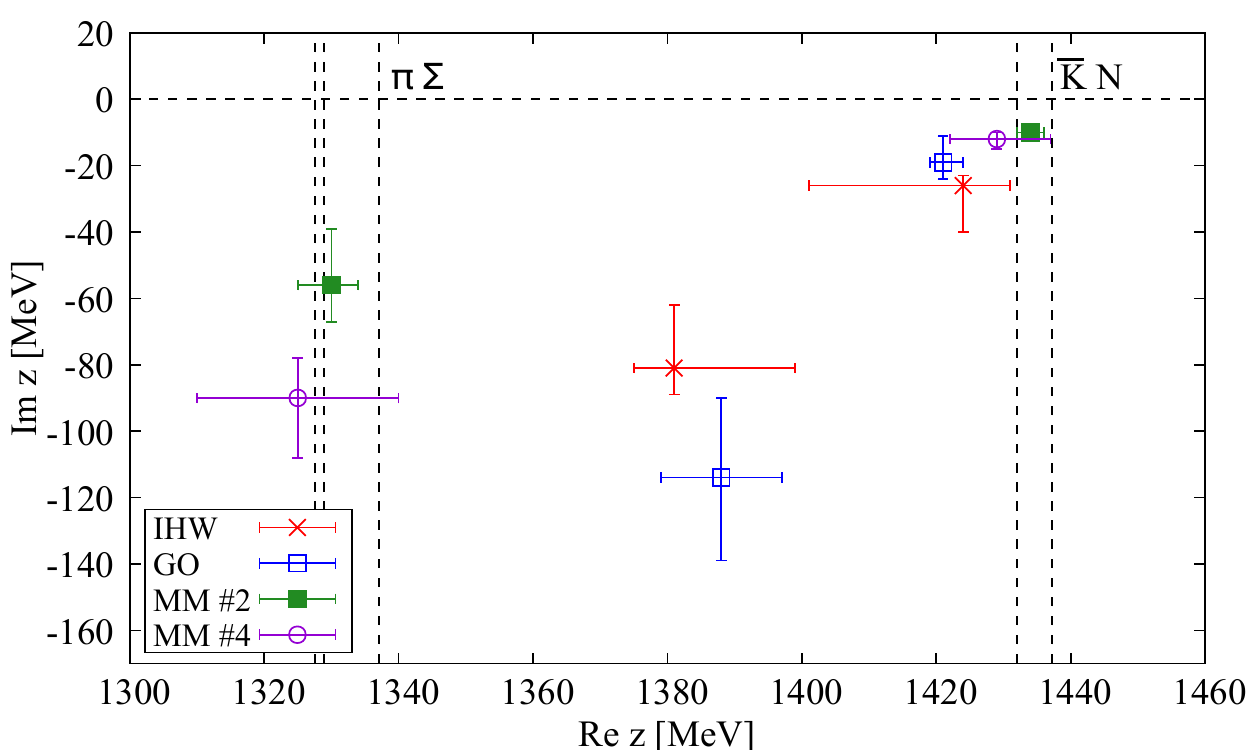}
  }
  \caption{Pole positions in the $\Lambda(1405)$ region from next-to-leading order chiral SU(3) dynamics including the SIDDHARTA constraint,
  IHW~\cite{Ikeda:2011pi,Ikeda:2012au}, GO~\cite{Guo:2012vv}, and MM~\cite{Mai:2014xna}.}
  \label{fig:comparison}
\end{figure}

The pole positions found in the $\Lambda(1405)$ energy region are summarized in Table~\ref{tab:comparison} and Fig.~\ref{fig:comparison}.\footnote{Here we show the best fit results in each analysis. In Refs.~\cite{Ikeda:2011pi,Ikeda:2012au}, the best fit result is obtained by the NLO approach. In Ref.~\cite{Guo:2012vv}, two results (Fit I and Fit II) are shown, but the Fit I is disfavored by the authors in view of recent studies~\cite{Ollerprivate}. In Ref.~\cite{Mai:2014xna}, the consistency check with the $\pi\Sigma$ spectra allows solutions \#2 and \#4, out of eight solutions found by analyzing the $K^{-}p$ scattering data.} Qualitatively, all the state-of-the-art studies~\cite{Ikeda:2011pi,Ikeda:2012au,Guo:2012vv,Mai:2014xna} found two poles in this energy region: one pole near the $\bar{K}N$ threshold (pole 1) and a second one at lower energy (pole 2). It should be emphasized that the two poles emerge as a consequence of the systematic fitting; their appearance is not enforced by hand. The position of pole 1 is consistently found in a narrow region above 1420 MeV. There is consensus about this feature of the $\Lambda(1405)$ in all chiral SU(3) dynamics analyses constrained by the $K^{-}p$ scattering and SIDDHARTA data. 

\begin{table}[bt]	
\caption{
Comparison of the pole positions in the $\Lambda(1405)$ region from next-to-leading order chiral SU(3) dynamics
including the SIDDHARTA constraint. \label{tab:comparison}
}
\begin{center}
\begin{tabular}{lll}
\hline
approach & pole 1 (MeV) & pole 2 (MeV) \\ \hline
Refs.~\cite{Ikeda:2011pi,Ikeda:2012au} NLO 
  & $1424^{+7}_{-23}- i 26^{+3}_{-14}$
  &  $1381^{+18}_{-6}- i 81^{+19}_{-8}$ \\
Ref.~\cite{Guo:2012vv} Fit II 
  & $1421^{+3}_{-2}- i 19^{+8}_{-5}$ 
  & $1388^{+9}_{-9}- i 114^{+24}_{-25}$ \\
Ref.~\cite{Mai:2014xna} solution \#2
  & $1434^{+2}_{-2} - i  10^{+2}_{-1}$ 
  & $1330^{+4~}_{-5~} - i  56^{+17}_{-11}$\\
Ref.~\cite{Mai:2014xna} solution \#4
  & $1429^{+8}_{-7} - i  12^{+2}_{-3}$ 
  & $1325^{+15}_{-15} - i  90^{+12}_{-18}$\\
\hline
\end{tabular}
  \end{center}
\end{table}

As can be seen in Fig.~\ref{fig:comparison} the position of pole 2 is subject to some ambiguities. This is because the main sources of the experimental data accumulated at and above the $\bar{K}N$ threshold. While the amplitude around the $\bar{K}N$ threshold is well determined, the constraint on the region far from the threshold is not very strong. To sharpen the description of the subthreshold $\bar{K}N$ amplitude, one should include the $\pi\Sigma$ invariant mass spectra in the fitting procedure. In fact, it turns out in Ref.~\cite{Mai:2014xna} that the consistency check with the $\pi\Sigma$ spectra is important to exclude some unphysical solutions. In the next section, we discuss the importance of precise empirical $\pi\Sigma$ spectra for the study of the $\Lambda(1405)$.

\section{$\Lambda(1405)$ and $\pi\Sigma$ spectra}

The $\Lambda(1405)$ decays exclusively into the $\pi\Sigma$ channel. Traditionally, the basic information about the $\Lambda(1405)$ came from the $\pi^{-}\Sigma^{+}$ spectrum measured in the $K^{-}p\to \Lambda(1405)\pi^{+}\pi^{-}\to \pi^{-}\Sigma^{+}\pi^{+}\pi^{-}$ reaction~\cite{Hemingway:1985pz}. The charged $\pi\Sigma$ state is, however, not the ideal channel to study the $\Lambda(1405)$. As pointed out in Ref.~\cite{Nacher:1998mi}, the isospin decomposition of the $\pi\Sigma$ spectrum in the charge basis is given by 
\begin{align}
    \frac{d\sigma(\pi^{+}\Sigma^{-})}{dM_{\rm inv}}
    &\propto \frac{1}{3}|T^{(0)}|^2+\frac{1}{2}|T^{(1)}|^2
    +\frac{2}{\sqrt{6}}\text{Re } (T^{(0)}T^{(1)*}) ,
    \label{eq:pipSigmam} \\
    \frac{d\sigma(\pi^{-}\Sigma^{+})}{dM_{\rm inv}}
    &\propto \frac{1}{3}|T^{(0)}|^2+\frac{1}{2}|T^{(1)}|^2
    -\frac{2}{\sqrt{6}}\text{Re } (T^{(0)}T^{(1)*}) ,
    \label{eq:pimSigmap} \\
    \frac{d\sigma(\pi^{0}\Sigma^{0})}{dM_{\rm inv}}
    &\propto \frac{1}{3}|T^{(0)}|^2  ,
    \label{eq:pi0Sigma0}
\end{align}
where $T^{(I)}$ is the meson-baryon scattering amplitude with isospin $I$ and the small $I=2$ component is neglected.\footnote{Strictly speaking, the relations~\eqref{eq:pipSigmam}, \eqref{eq:pimSigmap}, and \eqref{eq:pi0Sigma0} are obtained by assuming the same initial state and final state. In general, the initial state can be a more complicated combination as shown in Eq.~\eqref{eq:photopro}. Even in this case, the $I=1$ component and the interference terms generally appear in the charged $(\pi^\pm\Sigma^\mp)$ final states.} It is clear that the charged $\pi\Sigma$ spectra are proportional not only to the $I=0$ component but also to the $I=1$ contribution $|T^{(1)}|^2$ and the interference term $\text{Re } (T^{(0)}T^{(1)*})$, even in the isospin symmetric limit. These contaminations may disturb the signal of the $\Lambda(1405)$ in the $T^{(0)}$ amplitude.

The effect of the interference is indeed observed in recent experiments. In the photoproduction experiment $\gamma p\to K^{+}\Lambda(1405)\to K^{+}\pi^{\pm}\Sigma^{\mp}$ performed by the LEPS collaboration~\cite{Niiyama:2008rt}, different line shapes of the $\pi^{+}\Sigma^{-}$ and $\pi^{-}\Sigma^{+}$ modes are observed. This observation is elaborated in the comprehensive high-precision studies of the photoproduction by the CLAS collaboration~\cite{Moriya:2013eb,Moriya:2013hwg,Moriya:2014kpv}. They measured the spectra in all three charge combinations ($\pi^{+}\Sigma^{-}$, $\pi^{-}\Sigma^{+}$, and $\pi^{0}\Sigma^{0}$) at various initial photon energies ranging from 1.95 GeV to 2.85 GeV~\cite{Moriya:2013eb}, investigated the angular dependence~\cite{Moriya:2013hwg}, and even determine the spin and parity of the $\Lambda(1405)$ experimentally~\cite{Moriya:2014kpv}. Another important analysis has recently been performed in a different reaction, $pp\to K^{+}p\pi^{\pm}\Sigma^{\mp}$, by the HADES collaboration~\cite{Agakishiev:2012xk}. In all cases, the isospin interference effect in the charged $\pi\Sigma$ spectrum is clearly observed. An important achievement in these new experiments in comparison with the old measurement~\cite{Hemingway:1985pz} is the determination of the absolute values of the differential cross sections. This enables us to compare the results with the theoretical studies in a quantitative manner. Given the uncertainties in the subthreshold $\bar{K}N$ amplitude discussed in the previous section, it is of basic importance to use the different $\pi\Sigma$ spectra in order to constrain the subthreshold $\bar{K}N$ amplitude. In fact, the CLAS photoproduction data is used for this purpose in Refs.~\cite{Mai:2014xna,Roca:2013av,Roca:2013cca}. It was shown in Refs.~\cite{Roca:2013av,Roca:2013cca} that the photoproduction data alone can already provide the two poles of the $\Lambda(1405)$.

\begin{figure}[tb]
  \centerline{
  \includegraphics[width=5cm,bb=0 0 350 203]{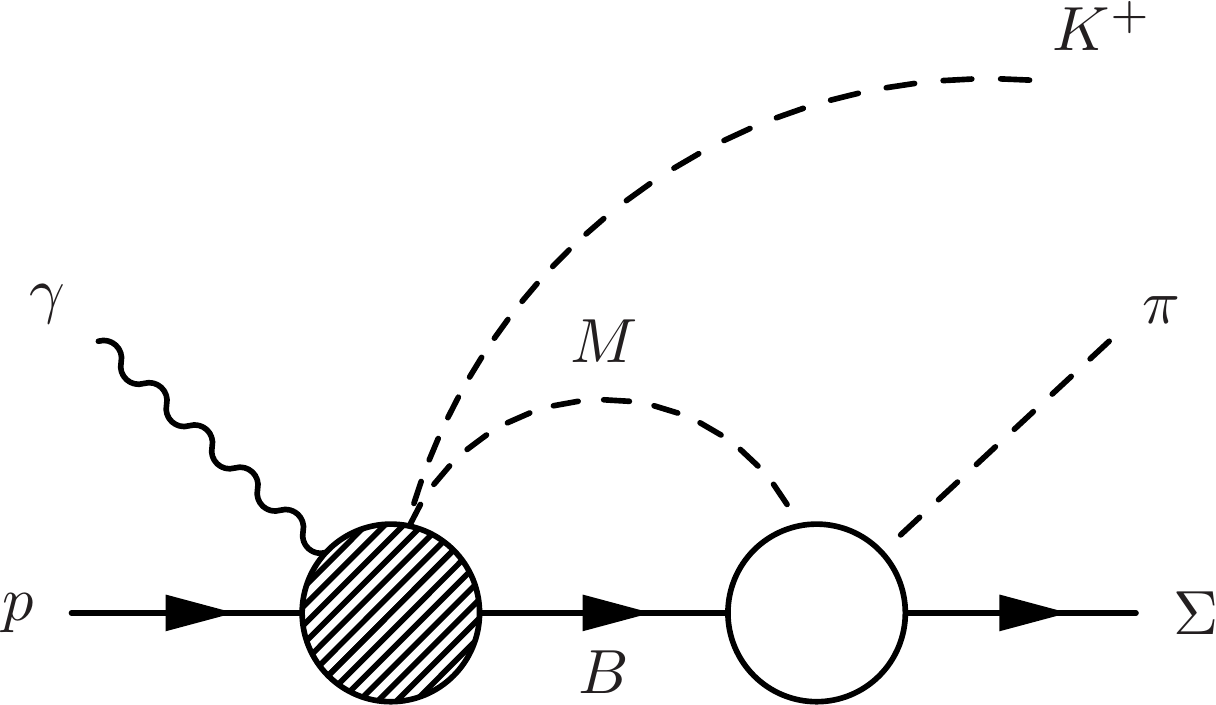}
  }
  \caption{One of the Feynman diagrams for the photoproduction $\gamma p\to K^{+}\pi\Sigma$. The shaded blob represents the initial state interaction $C_{MB}$ and the open blob stands for the meson-baryon scattering amplitude $T_{MB\to \pi\Sigma}(M_{\rm inv})$.}
  \label{fig:photopro}
\end{figure}

It should however be noted that the $\pi\Sigma$ spectra can only be obtained in production experiments, not in elastic scattering. This implies a somewhat involved procedure to extract the relevant two-body meson-baryon amplitudes from the observed data. Consider photoproduction as an example. In the $\gamma p\to K^{+}\pi\Sigma$ process, there is a contribution from the Feynman diagram shown in Fig.~\ref{fig:photopro}. The invariant mass distribution is calculated as 
\begin{align}
    \frac{d\sigma}{dM_{\rm inv}}
    &\propto
    \left|\sum_{MB}C_{MB}G_{MB}(M_{\rm inv})
    T_{MB \pi\Sigma}(M_{\rm inv})\right|^{2}
    ,
    \label{eq:photopro}
\end{align}
where $C_{MB}$ denotes the initial state interaction, and $T_{MB \pi\Sigma}(M_{\rm inv})$ is the two-body scattering amplitude from $MB$ to $\pi\Sigma$ in the final state interaction at the invariant mass $M_{\rm inv}$ of the $\pi\Sigma$ system. Equation~\eqref{eq:photopro} implies that, in order to extract $T_{MB\pi\Sigma}(M_{\rm inv})$ from the experimental differential cross section $d\sigma/dM_{\rm inv}$, we need to calculate the initial state interaction, $C_{MB}$, by introducing a theoretical model. In general, $C_{MB}$ depends on the total energy of the $\gamma p$ system and the scattering angle of the final $K^{+}$. Thus the $\pi\Sigma$ spectrum does not provide a direct constraint on the two-body amplitude, in contrast to the kaonic hydrogen measurement~\eqref{eq:shiftwidth} which directly determines $T_{K^{-}pK^{-}p}(m_{K^{-}}+M_{p})$. 

Moreover, the diagram in Fig.~\ref{fig:photopro} is just one out of many processes that contribute to $\gamma p\to K^{+}\pi\Sigma$. For instance, there can be final state interactions in the $K^{+}\pi$ and $K^{+}\Sigma$ systems. The final state interactions can take place subsequently (for instance, the $K^{+}\pi$ interacts after the $\pi\Sigma$ interaction). Contributions other than those of the diagram in Fig.~\ref{fig:photopro} are not always written in the generic form of Eq.~\eqref{eq:photopro}. From the theoretical viewpoint, the reaction mechanism should be studied in each process as accurately as possible, in order to extract the information of the two-body amplitude from the observed spectrum. 

With these caveats in mind, in the following, we introduce two recent studies of $\pi\Sigma$ spectra in the region of the $\Lambda(1405)$.

\subsection{$K^{-}d\to \pi\Sigma n$ reaction: full three-body dynamics}

We first consider the $K^{-}d\to \pi\Sigma n$ reaction, which is currently under investigation by the J-PARC E31 collaboration. With an incoming 1 GeV $K^{-}$ beam, the missing mass spectrum of the forward-going neutron will be measured. All three charge combinations of the final $\pi\Sigma$ states will be separated according to their decay products detected by the Cylindrical Detector System (CDS). 
On the theoretical side this reaction was previously studied using approximate two-step approaches in Refs.~\cite{Jido:2009jf,Jido:2010rx,Miyagawa:2012xz,Jido:2012cy,YamagataSekihara:2012yv}. The two-step approach corresponds to a truncation of the full three-body amplitude of the final state interactions which can be calculated by solving Faddeev-type equations. Three-body calculations of this reaction were recently performed, but for low-energy kinematics~\cite{Revai:2012fx,Shevchenko:2014uva}. 

\begin{figure}[tb]
  \centerline{
  \includegraphics[width=7cm,bb=0 0 360 252]{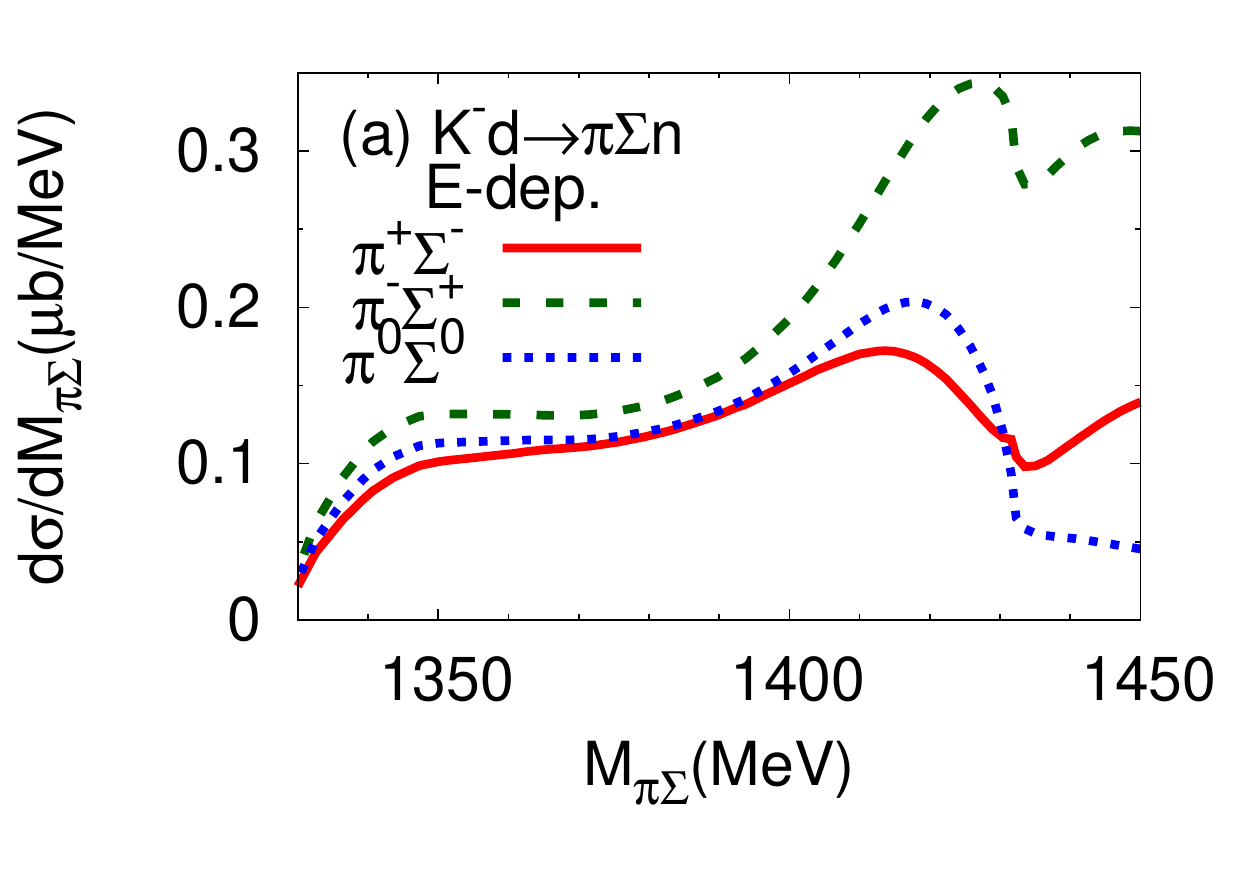}
  \includegraphics[width=7cm,bb=0 0 360 252]{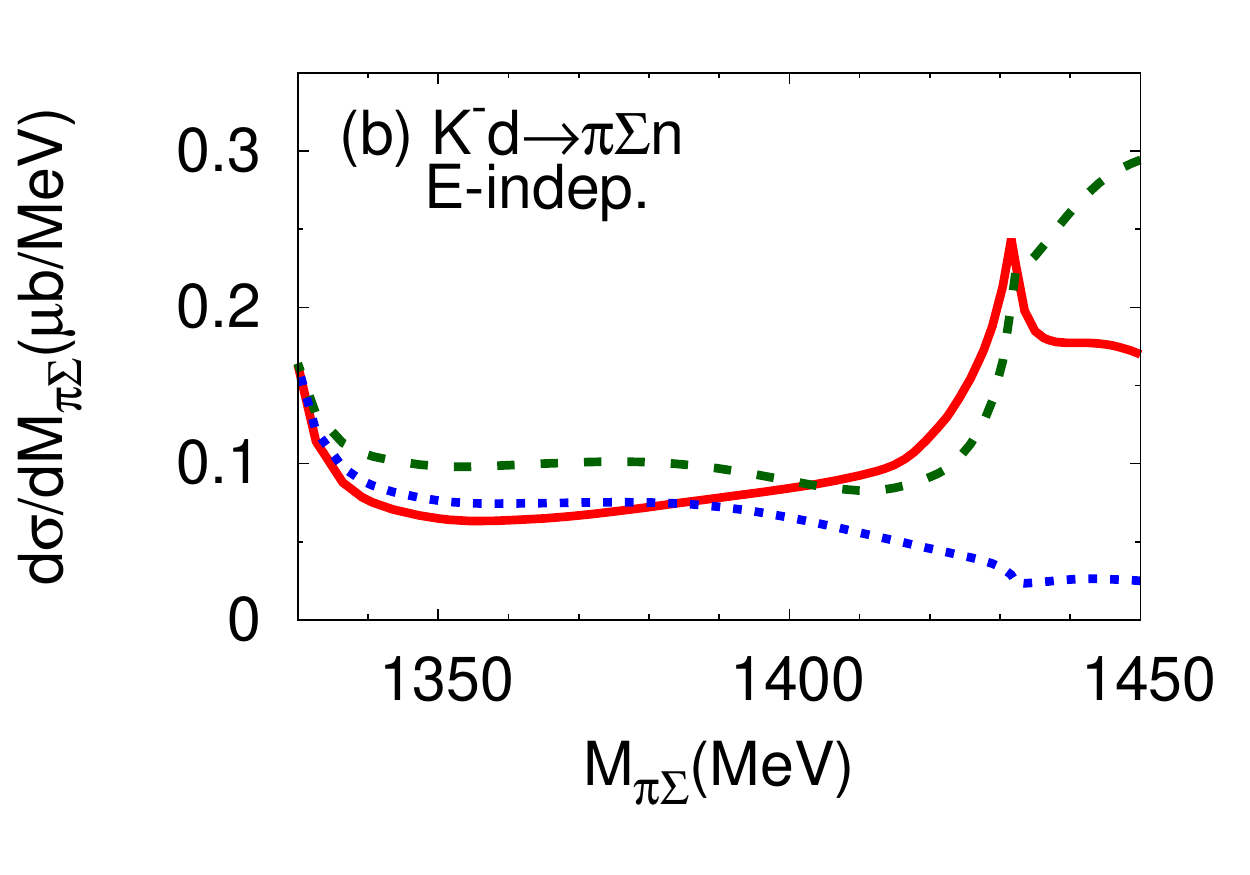}
  }
  \caption{Differential cross sections 
 {$d\sigma/dM_{\pi\Sigma}$} 
 for $K^-  d\rightarrow \pi\Sigma n$.
 The initial kaon momentum is set to $p_{\rm lab}= 1$~GeV.
 Panel~(a): E-dep.~model; Panel~(b) E-indep.~model.
 Solid curves: $\pi^+\Sigma^-n$;
dashed curves: $\pi^-\Sigma^+n$;
dotted curves: $\pi^0\Sigma^0n$ in the final
state, respectively.}
  \label{fig:Kd}
\end{figure}%

In Ref.~\cite{Ohnishi:2015iaq}, the $K^{-}d\to \pi\Sigma n$ reaction is studied performing a full three-body calculation with kinematical conditions adapted to the J-PARC experiment. The coupled-channels Alt-Grassberger-Sandhas (AGS) equations for the $\bar{K}NN$-$\pi YN$ system are solved with relativistic kinematics. For technical reasons the NLO chiral SU(3) amplitude cannot be directly used in the AGS equations, hence we employ the $\bar{K}N$ interaction based on the Weinberg-Tomozawa term which is sufficient for this purpose. To see the sensitivity of the $\pi\Sigma$ spectrum to the two-body interaction, we examine two interaction models with different subthreshold behaviors: an "E-dep." model with energy-dependent interactions derived from the chiral SU(3) effective Lagrangian, and an "E-indep." model that approximates these interactions by constants fixed at the $\bar{K}N$ threshold~\cite{Ikeda:2007nz,Ikeda:2008ub,Ikeda:2010tk,Ikeda:2011dx,Ohnishi:2013rix}. The cutoff parameters in both models are adjusted to reproduce the $K^{-}p$ scattering data. The $K^{-}p$ scattering length in the E-dep.~model is consistent with the SIDDHARTA measurements. The subthreshold amplitude with two poles ($1428.8-i\,15.3$ MeV and $1344.0-i\,49.0$ MeV) behaves similarly as in NLO chiral SU(3) dynamics~\cite{Ohnishi:2015iaq}. The E-indep.~model does not reproduce the SIDDHARTA data and only a single pole is found in the amplitude at $1405.8-i\,25.2$ MeV.

\begin{figure}[tb]
  \centerline{
  \includegraphics[width=7cm,bb=0 0 360 252]{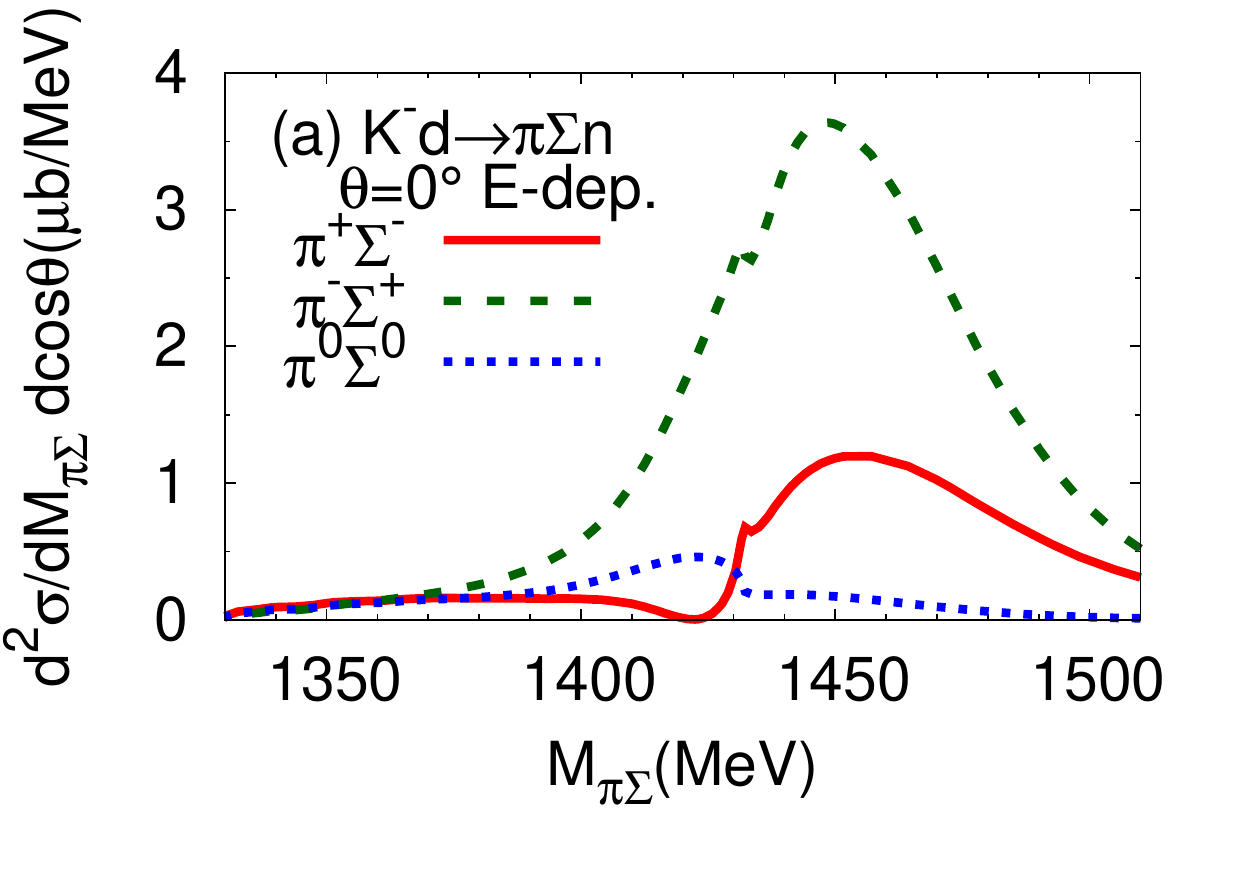}
  \includegraphics[width=7cm,bb=0 0 360 252]{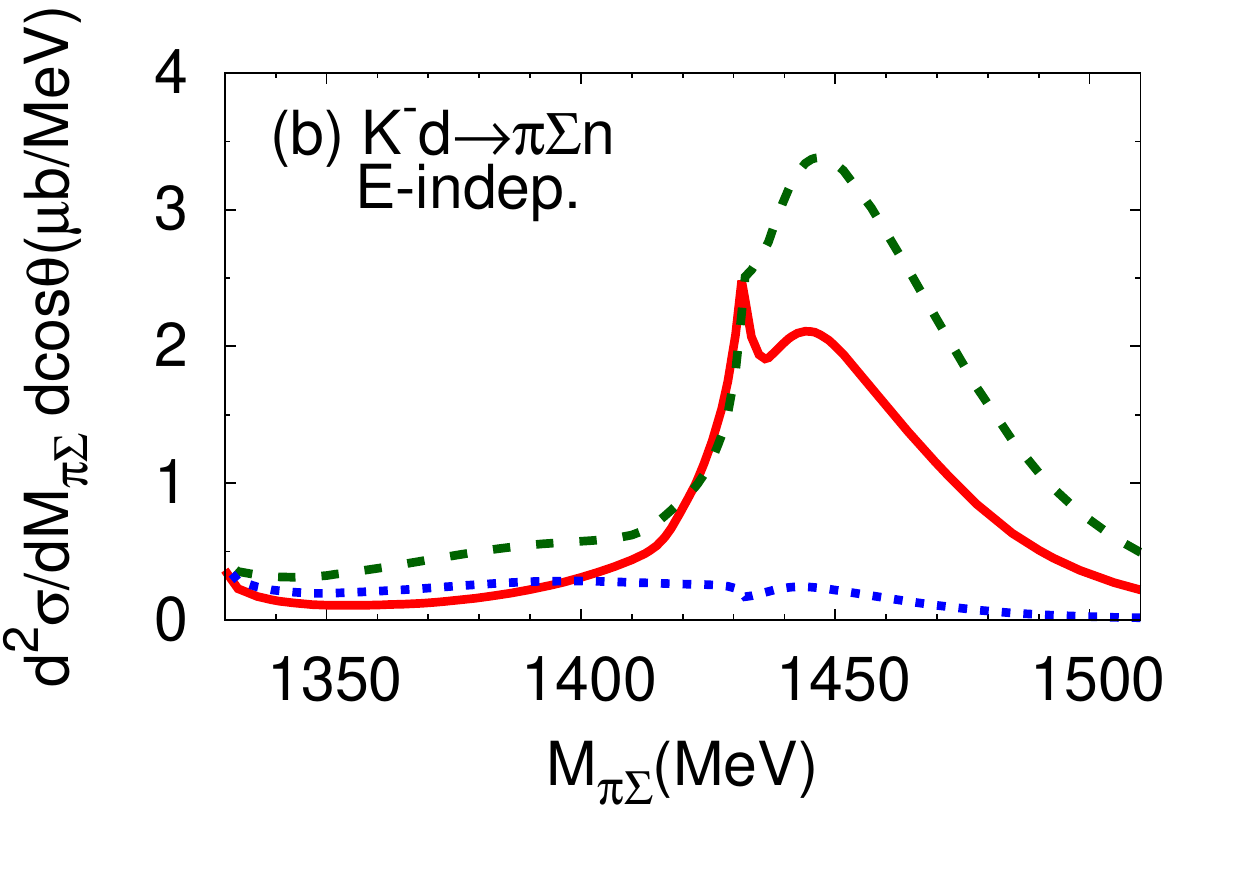}
  }
  \caption{Double differential cross sections 
 {$d^2\sigma/dM_{\pi\Sigma}d\cos \theta$} 
 for $K^- d\rightarrow \pi\Sigma n$ with the neutron emitted in
 forward direction, $\theta=0^\circ$.
 Panel~(a): the E-dep.~model; Panel~(b) the E-indep.~model.
 Solid curves: $\pi^+\Sigma^-n$;
dashed curves: $\pi^-\Sigma^+n$;
dotted curves: $\pi^0\Sigma^0n$ in the final
state, respectively.
 The incident $K^-$ momentum is $p_{\rm lab}= 1$~GeV.}
  \label{fig:Kd0deg}
\end{figure}%

Differential cross sections $d\sigma/dM_{\pi\Sigma}$ for the $K^- d\rightarrow \pi\Sigma n$ reaction calculated with the E-dep.~model are shown in Fig.~\ref{fig:Kd} (a) and for the E-indep.~model in Fig.~\ref{fig:Kd} (b). The initial energy is chosen to correspond to that of the J-PARC experiment. While the resonant structure of the $\Lambda(1405)$ is visible in the E-dep.~model, this structure is flattened in the E-indep.~model by competing contributions even though the $\Lambda(1405)$ pole exists in the amplitude. It is evident that the $\pi\Sigma$ spectra are sensitive to the detailed subthreshold behavior of the $\bar{K}N$ amplitude. In both cases the isospin interference effects are observed as differences between separate $\pi\Sigma$ charge modes. Figure~\ref{fig:Kd0deg} shows the double differential cross sections, $d^2\sigma/dM_{\pi\Sigma}\,d\cos \theta$, with the emitted neutron going in forward direction, $\theta=0^{\circ}$, as in the actual experiment. The $\Lambda(1405)$ maximum in the E-dep.~model is not visible any more, with the exception of the $\pi^{0}\Sigma^{0}$ channel. This is caused by the strong isospin interference mechanism. In other words, the effect of the $I=1$ amplitude is enhanced in the forward-neutron spectra. We also note a pronounced bump structure above the $\bar{K}N$ threshold (around $M_{\pi\Sigma}\sim 1450$ MeV) in both the E-dep.~and E-indep.~models, which is attributed to  subtle three-body coupled-channels mechanisms~\cite{Ohnishi:2015iaq}. The experimental data by the J-PARC E31 collaboration will provide new information on the subthreshold $\bar{K}N$ amplitude through the comparison with these predicted spectra.

\subsection{Nonleptonic weak decay of $\Lambda_{c}\to \pi^{+}MB$}

Recently, the weak decay of heavy hadrons is found to be a useful tool to study resonances in hadronic final state interactions (see Ref.~\cite{Oset:2016lyh} for a recent review). A striking finding in 2015 has been the observation of the $J/\psi p$ resonance in the $\Lambda_{b}\to J/\psi K^{-}p$ decay by the LHCb collaboration~\cite{Aaij:2015tga}. Apart from the exotic pentaquark candidate, several excited $\Lambda^{*}$ states are observed in the $K^{-}p$ spectrum. This opens the possibility of studying $S=-1$ baryon resonances in this decay process, as theoretically discussed in Refs.~\cite{Roca:2015tea,Roca:2015dva,Feijoo:2015cca}. Here we focus on the decay of the charmed baryon $\Lambda_{c}\to \pi^{+}MB$ with $MB=\bar{K}N$ and $\pi\Sigma$. Recently, these decay channels have been studied in detail by the Belle~\cite{Zupanc:2013iki} and BESIII~\cite{Ablikim:2015flg} collaborations. It is also suggested theoretically to use the cusp effect in the $\pi\pi\Sigma$ decay for the determination of the $\pi\Sigma$ scattering length~\cite{Hyodo:2011js}.

The effect of the $\Lambda(1405)$ in the $\pi\Sigma$ spectra of the $\Lambda_{c}$ decay is studied in Ref.~\cite{Miyahara:2015cja}. For the kinematics of the $\Lambda(1405)$ production, the emitted $\pi^{+}$ has a large momentum. In this case, from the combined viewpoints of the Cabbibo-Kobayashi-Maskawa matrix, color suppression and diquark correlations in the $\Lambda_{c}$, the dominant contribution of the initial weak decay process generates a meson-baryon pair in the combination
\begin{align}
    \ket{MB} 
    &= \ket{K^-p} + \ket{\bar{K}^0n} 
    -\frac{\sqrt{2}}{3}\ket{\eta\Lambda}.  \label{eq:hadronstate} 
\end{align}
An important observation is that this combination is pure isospin $I=0$. We expect that the strong final state interaction $MB\to \pi\Sigma$ will not be affected by isospin interference because of the dominance of this mechanism. Hence the $\Lambda_{c}$ decay process acts as an isospin filter for the $\pi\Sigma$ spectra.

\begin{figure}[tb]
  \centerline{
  \includegraphics[width=8cm,bb=0 0 659 464]{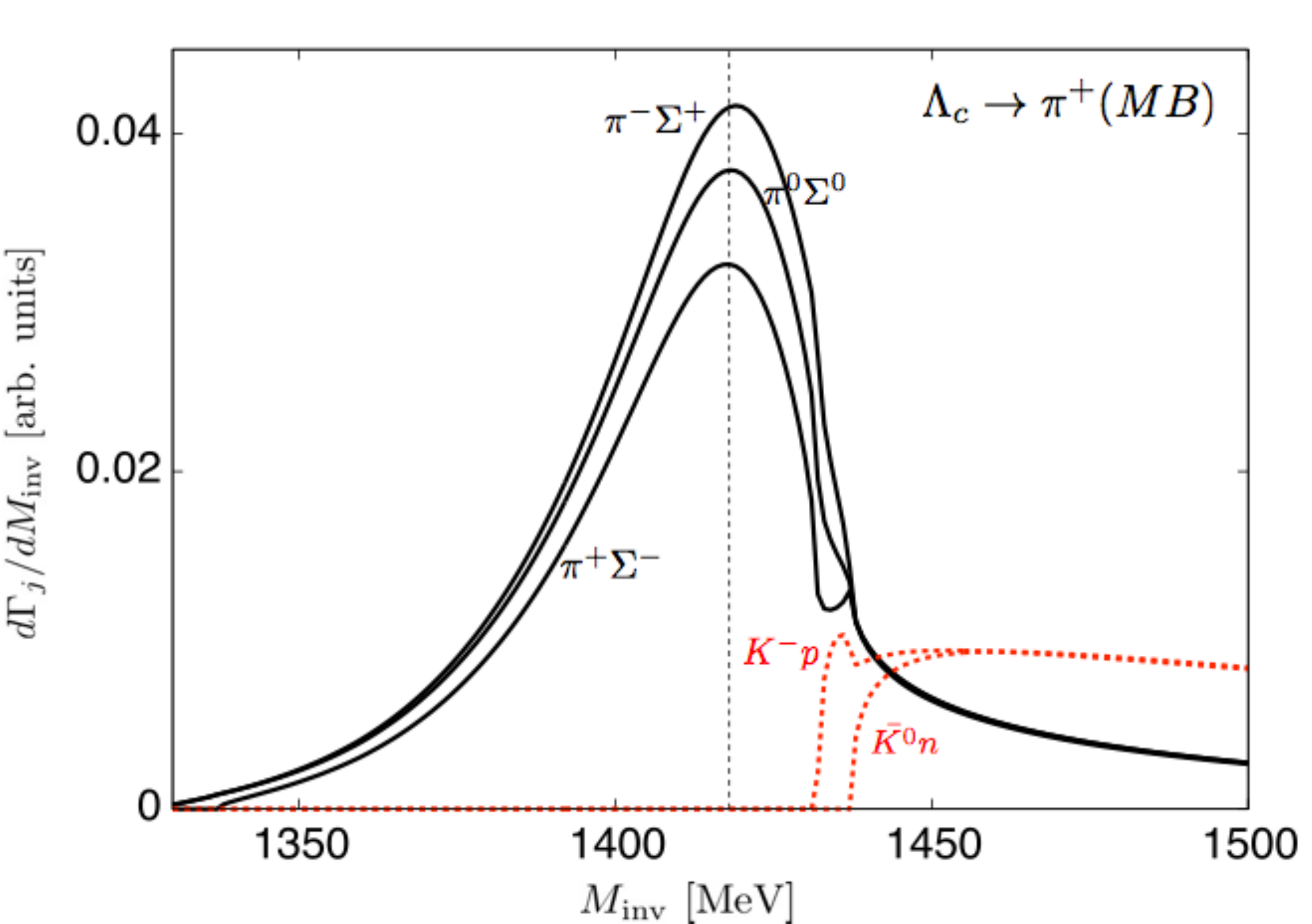}
  }
  \caption{Invariant mass distribution of the decay $\Lambda_c^+\to \pi^+ MB$ near the $\bar{K}N$ threshold with the meson-baryon scattering amplitude in Refs.~\cite{Ikeda:2011pi,Ikeda:2012au}. The solid line represents the spectrum for $\pi\Sigma$ channels and the dashed line for $\bar{K}N$ channels. Vertical dashed line indicates $M_{\rm inv}=1418$ MeV.}
  \label{fig:Lambdac}
\end{figure}

Figure~\ref{fig:Lambdac} shows the predicted $\pi\Sigma$ and $\bar{K}N$ spectra in the $\Lambda_{c}\to \pi^{+}MB$ decay. The final state interaction is accounted for by NLO chiral SU(3) dynamics as in Refs.~\cite{Ikeda:2011pi,Ikeda:2012au}. In contrast to the previously discussed processes, all charge combinations of the $\pi\Sigma$ states show a maximum at the same energy, as indicated by the vertical dashed line in Fig.~\ref{fig:Lambdac}. This is a consequence of the isospin filter mechanism of the $\Lambda_{c}$ decay. Small deviations are caused by the isospin breaking effect in the meson-baryon amplitude with physical hadron masses~\cite{Ikeda:2011pi,Ikeda:2012au}. In addition the peak position is found around 1418 MeV, because the initial configuration~\eqref{eq:hadronstate} does not involve $\pi\Sigma$ states and puts more weight on the higher energy pole of the $\Lambda(1405)$ coupled-channels system. The suppression of the isospin interference effect in the $\pi\Sigma$ spectra is advantageous for the experimental observation as the charged final states are in general easy to detect.

\section{Internal structure of $\Lambda(1405)$}

Finally, we discuss the structure of the $\Lambda(1405)$, focusing on the $\bar{K}N$ molecular component. It is suggestive that recent lattice QCD results point out evidence for the molecular $\bar{K}N$ structure of the $\Lambda(1405)$~\cite{Hall:2014uca}. The $\bar{K}N$ molecular picture of the $\Lambda(1405)$ is also the basis for calculations of few-body $\bar{K}$ nuclei~\cite{Shevchenko:2006xy,Shevchenko:2007ke,Yamazaki:2007cs,Ikeda:2007nz,Dote:2008in,Dote:2008hw,Ikeda:2010tk,Bayar:2011qj,Barnea:2012qa,Bayar:2012rk,Bayar:2013lxa,Revai:2014twa}. 

In the framework of chiral SU(3) dynamics, one may naively think that the $\bar{K}N$ molecular picture of the $\Lambda(1405)$ is somehow built in from the beginning, because the model space does not explicitly include the $qqq$ component of the $\Lambda(1405)$. This simple expectation is, however, not always realized, as demonstrated for example in the meson sector. In the $N_{c}$ scaling analysis of scalar and vector mesons produced in meson-meson scattering~\cite{Pelaez:2003dy,Pelaez:2006nj}, the nature of the $\rho$ meson is found to be that of a $\bar{q}q$ state, even though the model space is built only with pseudoscalar mesons\footnote{A similar $N_{c}$ scaling study of the $\Lambda(1405)$ shows that both of the poles exhibit the non-$qqq$ behavior~\cite{Hyodo:2007np,Roca:2008kr}. This is consistent with the $\bar{K}N$ molecular picture, though this analysis does not specify this explicit component.}. In general, contributions other than those included in the model space are hidden in the low-energy constants of the chiral effective Lagrangian~\cite{Ecker:1988te} at higher order or in the renormalization procedure~\cite{Hyodo:2008xr}. The $\rho$ meson is actually generated mostly from the NLO terms of the chiral Lagrangian, which have encoded information of the vector mesons. Hence, the $\bar{K}N$ molecular picture of the $\Lambda(1405)$ should be examined also in chiral SU(3) dynamics.

Here we report on recent studies of the structure of the $\Lambda(1405)$ from the viewpoint of the $\bar{K}N$ wavefunction and compositeness. As discussed in section~\ref{sec:polestructure}, there are two poles in the relevant energy region. At each pole, the wavefunction and the compositeness can be calculated, reflecting the nature of the corresponding eigenstate. It is shown that the position of the higher energy pole has small systematic uncertainty, and it should have strong effects on the $\bar{K}N$ amplitude around the threshold because of its location. We thus mainly concentrate on the structure of this higher energy eigenstate.

\subsection{Realistic $\bar{K}N$ potential and wavefunction of $\Lambda(1405)$}

The wavefunction in coordinate space reflects the spatial structure of the eigenstate. It is shown in Ref.~\cite{Hyodo:2007jq} that a single-channel effective $\bar{K}N$ potential can be constructed so as to reproduce the equivalent $\bar{K}N$ scattering amplitude derived from chiral SU(3) dynamics. The original aim of Ref.~\cite{Hyodo:2007jq} was the application of the potential to few-nucleon systems with an antikaon~\cite{Dote:2008in,Dote:2008hw}. In the two-body  $\bar{K}N$ sector the structure of the $\Lambda(1405)$ can be studied examining the wavefunction in this channel.

For a quantitative discussion of the $\Lambda(1405)$ structure, one must start from a reliable $\bar{K}N$ potential. A suitable potential for this purpose is constructed in Ref.~\cite{Miyahara:2015bya} based on the scattering amplitude derived from NLO chiral SU(3) dynamics~\cite{Ikeda:2011pi,Ikeda:2012au}. By construction, this potential $U(r,E)$ is energy dependent and reproduces the experimental data with an accuracy of $\chi^{2}/\rm{dof}\sim 1$. The coordinate space wavefunction $\psi(r)$ can be obtained by solving the Schr\"odinger equation at the energy eigenvalue.

\begin{figure}[tb]
  \centerline{
  \includegraphics[width=8cm,bb=0 0 764 519]{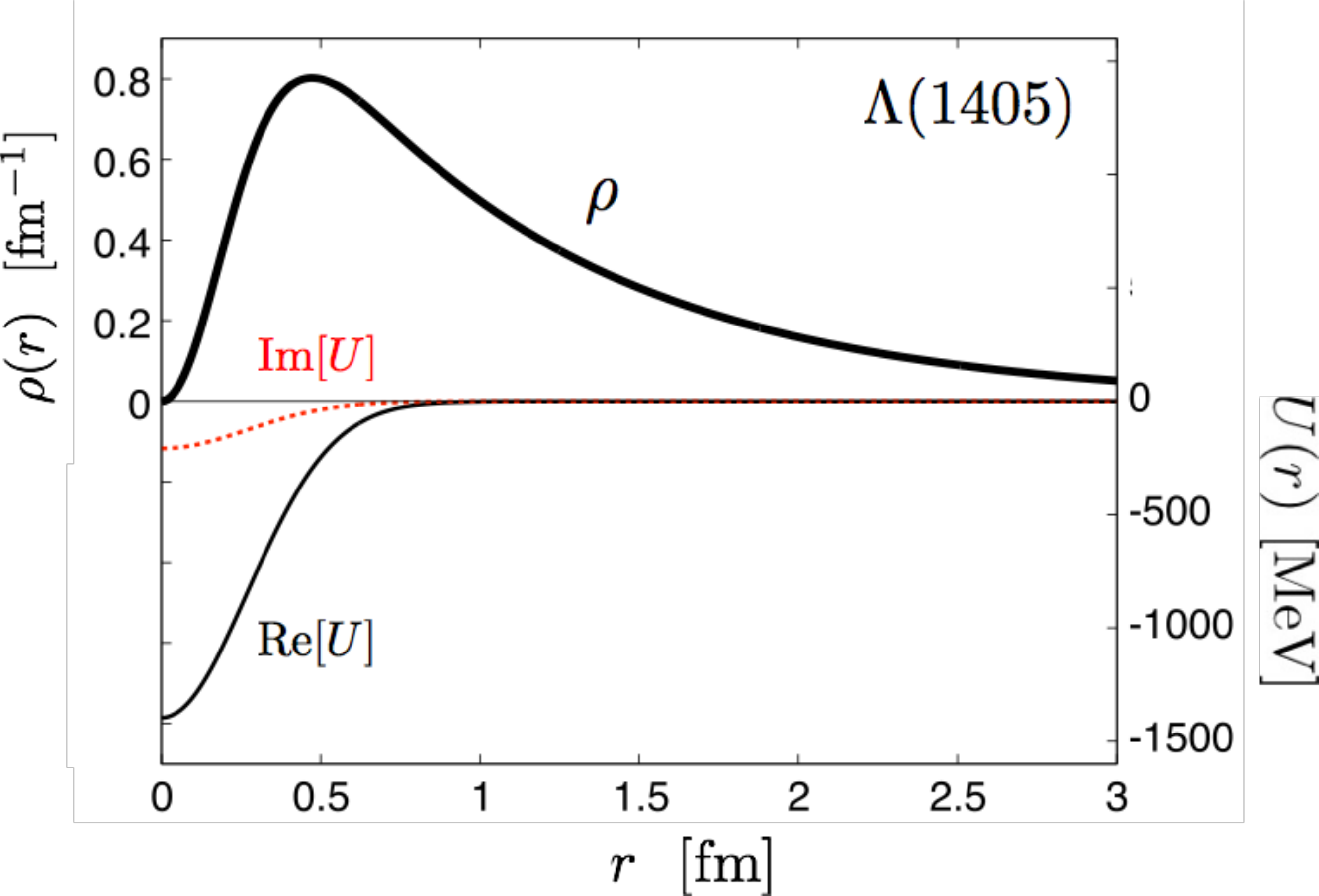}
  }
  \caption{$\bar{K}N$ density distribution $\rho(r) = r^{2}|\psi(r)|^{2}$ (thick solid line), the real part (thin solid line) and the imaginary part (dotted line) of the $\bar{K}N$ potential $(I=0)$ at the energy of the higher pole of the $\Lambda(1405)$.}
  \label{fig:potential}
\end{figure}

The real and imaginary parts of the $\bar{K}N$ potential $U(r,E)$ at the energy of the higher pole are shown in Fig.~\ref{fig:potential}. The real part of the potential is negative, indicating the attractive nature of the interaction. The imaginary part represents the absorption by the transition to the $\pi\Sigma$ channel. In Fig.~\ref{fig:potential}, the $r^2$-weighted density distribution $\rho(r) = r^{2}|\psi(r)|^{2}$ is also plotted. A substantial part of the $\bar{K}N$ density is located in the region at $r> 1$ fm, outside of the potential range. The mean  $\bar{K}N$ distance is obtained as\footnote{We note that the normalization of an unstable state should be done with caution. For instance, with the Gamow state normalization, the mean distance is obtained as a complex number. As discussed in the Appendix of Ref.~\cite{Miyahara:2015bya}, the definition of Eq.~\eqref{eq:distance} can be interpreted as the spatial extent, when the result is much larger than the potential range.}
\begin{align}
    \sqrt{\langle r^{2}\rangle}
    &= \left(\int d^{3}r \rho(r)\right)^{1/2}
    =1.44 \phantom{0}\rm{fm}.  \label{eq:distance} 
\end{align}
This spatial extension is significantly larger than typical hadron sizes, $r\lesssim 1$ fm. For comparison, the mean distance of the lower energy eigenstate associated with the second pole is 0.84 fm. The extended spatial structure of the higher energy eigenstate supports the $\bar{K}N$ molecular picture of the $\Lambda(1405)$. Similar results can be found in the studies with the Weinberg-Tomozawa approach~\cite{Dote:2008in,Sekihara:2008qk,YamagataSekihara:2010pj,Sekihara:2010uz,Sekihara:2012xp}.

\subsection{Compositeness of the $\Lambda(1405)$}

The wavefunction is not an observable: a unitary transformation of the wavefunction can be performed without changing observables. However, the long-distance behavior of the wavefunction of a weakly bound $s$-wave state is indeed related to characteristic observables. The large spatial size indicates a large scattering length. The low-energy behavior of a system with a very large scattering length is known to be universal~\cite{Braaten:2004rn}. The composite nature of a weakly bound $s$-wave state is determined by the scattering length and the binding energy~\cite{Weinberg:1965zz}. This model-independent weak-binding relation can now be regarded as a consequence of low-energy universality. Because such a model-independent relation is a powerful tool, the structure of hadrons are recently discussed from the viewpoint of compositeness~\cite{Baru:2003qq,Gamermann:2009uq,Hyodo:2011qc,Aceti:2012dd,Hyodo:2013iga,Hyodo:2013nka,Aceti:2014ala,Hyodo:2014bda,Sekihara:2014kya,Guo:2015daa}.

For the application to hadrons such as the $\Lambda(1405)$, the original relation in Ref.~\cite{Weinberg:1965zz} needs to be generalized to the case of unstable states. To this end, the nonrelativistic effective field theory (EFT) is adopted in Ref.~\cite{Kamiya:2015aea}. The EFT approach is a standard tool to study aspects of low-energy universality, and it is applicable to describe the low-energy behavior of any microscopic theory with a short-range interaction. It is demonstrated that the relation for the stable bound state in Ref.~\cite{Weinberg:1965zz} can be derived in the EFT approach, with a systematic assessment of correction terms. Moreover, a generalized relation can be obtained in the EFT approach, which was not possible in the original derivation using the completeness relation of the full Hamiltonian~\cite{Weinberg:1965zz}. This generalized relation reads
\begin{align}
    a_0
    &=  R \Biggl\{\frac{2X}{1+X} + {\mathcal O}
    \left(\left|\frac{R_{\mathrm{typ}}}{R}\right| \right) 
    + \sqrt{\frac{\mu^{\prime 3}}{\mu^{3}}} \mathcal{O} 
    \left( \left| \frac{l}{R} \right|^{3}\right) \Biggr\},
    \quad R=\frac{1}{\sqrt{-2\mu E_{QB}}} ,
    \label{eq:compositeness}
\end{align}
where $a_{0}$ is the scattering length,  $E_{QB}$ is the eigenenergy of the quasi-bound state, and $X$ is the compositeness. Here $\mu$ and $\mu^{\prime}$ are the reduced masses of the scattering and decay channels, respectively. The correction terms are estimated by the ratio of the length scales: $R$ is the length scale of the eigenstate, $R_{\rm typ}$ is the typical range of the two-body interaction of the microscopic theory, and $l=1/\sqrt{2\mu \nu}$ is the length scale associated with the threshold energy difference $\nu$. In the expansion of the scattering length $a_{0}$ in $1/R$, we find that the coefficient of the leading order term is given by the compositeness $X$. This means that the compositeness $X$ can be determined entirely by the observable quantities, $a_{0}$ and $E_{QB}$, when the magnitude of the eigenenergy is small enough to ignore higher order correction terms.

For a stable bound state, the compositeness $X$ is real and positive. It is interpreted as the probability of finding the two-body composite component of that state. For an unstable quasi-bound state, $X$ is given by a complex number, so we have to establish a prescription for its interpretation. In Ref.~\cite{Kamiya:2015aea}, three real quantities are introduced:
\begin{align}
    \tilde{X}
    &=  \frac{1-|1-X|+|X|}{2} ,\quad
    \tilde{Z}
    =  \frac{1+|1-X|-|X|}{2} , \quad
    U
    =|1-X|+|X|-1 .
\end{align}
It is shown that $U$ quantifies the deviation from the narrow width (bound state) limit, $\tilde{X}$ reduces to $X$ in the narrow width limit, and the real and positive $\tilde{X}$ and $\tilde{Z}$ satisfy $\tilde{X}+\tilde{Z}=1$. This means that $\tilde{X}$ can be interpreted as the probability of finding the composite component, when the uncertainty measure $U$ is small, while $\tilde{Z}$ represents the remaining components of the system such that the total probability adds up to unity.

We now evaluate the compositeness of the $\Lambda(1405)$. Currently, the most reliable determination of the pole energy and $\bar{K}N$ scattering length is provided by NLO chiral SU(3) dynamics. The compositeness of the  $\bar{K}N$ pole near threshold can be studied with small correction terms in Eq.~\eqref{eq:compositeness}. In Table~\ref{tab:Lambda} we summarize the energy eigenvalues of this higher-energy pole measured from the $\bar{K}N$ threshold, and the $\bar{K}N$ scattering lengths in the $I=0$ channel in Refs.~\cite{Ikeda:2011pi,Ikeda:2012au,Guo:2012vv,Mai:2014xna}. These eigenenergies satisfy $|R| \gtrsim 1.5\hspace{1ex}\mathrm{fm}$ so that the correction terms are found to be small, $|R_{\mathrm{typ}}/R|\lesssim 0.17$ and $|l/R|^3 \lesssim 0.14$ where $R_{\mathrm{typ}}$ and $l$ are estimated using $\rho$ meson exchange and the energy difference from the $\pi\Sigma$ threshold, respectively. Neglecting the small correction terms, we calculate the $\bar{K}N$ compositeness $X_{\bar{K}N}$, $\tilde{X}_{\bar{K}N}$, and the uncertainty in the interpretation $U$ as shown in Table~\ref{tab:Lambda}. We find indeed that the compositeness is close to unity with small uncertainty $U$. This indicates the dominance of the $\bar{K}N$ composite structure in the $\Lambda(1405)$ characteristic of a quasi-molecular state.  This result is consistent with other evaluations of the compositeness of the $\Lambda(1405)$ in Refs.~\cite{Sekihara:2012xp,Sekihara:2014kya,Garcia-Recio:2015jsa,Guo:2015daa}.

\begin{table}[bt]
		\caption{Compositeness of the $\Lambda (1405)$. Shown are the eigenenergy $E_{QB}$, $\bar{K}N(I=0)$ scattering length $a_{0}$, the $\bar{K}N$ compositeness $X_{\bar{K}N}$ and $\tilde{X}_{\bar{K}N}$, and the uncertainty of the interpretation  $U$.
		\label{tab:Lambda}
		}
  \begin{center}
		\begin{tabular}{llllll}
		\hline
             approach & $E_{QB}$ (MeV) & $a_0 $ (fm) & $X_{\bar{K}N}$ & $\tilde{X}_{\bar{K}N}$ & $U$  \\  \hline
			 Refs.~\cite{Ikeda:2011pi,Ikeda:2012au} NLO   & $-10-i26$ & $1.39 - i 0.85$ 
			 & $1.2+i0.1$ & $1.0$ & 0.5 \\ 
			 Ref.~\cite{Guo:2012vv} Fit II  & $-13-i20$ & $1.30-i0.85$ 
			 & $0.9-i0.2$ & $0.9$ & 0.1 \\
			 Ref.~\cite{Mai:2014xna} solution \#2 & $\phantom{-0}2-i10$ & $1.21-i1.47$ 
			 & $0.6+i0.0$ & $0.6$ & 0.0 \\ 
			 Ref.~\cite{Mai:2014xna} solution \#4 & $-\phantom{0}3-i12$ & $1.52-i1.85$ 
			 & $1.0+i0.5$ & $0.8$ & 0.6\\ 
			 \hline
		\end{tabular} 
  \end{center}
\end{table}

\section{Conclusions}

Chiral SU(3) dynamics is established as a powerful and systematic approach to study the $\Lambda(1405)$ and the $\bar{K}N$ interaction. The conclusions of this paper are summarized as answers to the questions raised in the introduction, as follows: 

\begin{enumerate}

\item[1.] The $\bar{K}N$ amplitude featuring the $\Lambda(1405)$ is described by next-to-leading order chiral SU(3) dynamics with an accuracy of $\chi^{2}$/d.o.f $\sim 1$. Two poles are found in the $\Lambda(1405)$ region when the calculations are confronted with the available empirical data base. The position of the higher energy pole, the one located just below the $\bar{K}N$ threshold, is well constrained by the precise determination of the $K^{-}p$ scattering length. 

\item[2.] There remain uncertainties in the energy region further below the $\bar{K}N$ threshold, including the position of the lower energy pole of the $\Lambda(1405)$. Detailed theoretical studies and precise experimental data of $\pi\Sigma$ mass spectra will provide further important constraints at the next stage. Full three-body calculation of the $K^{-}d\to \pi\Sigma n$ reaction point out the sensitivity of the $\pi\Sigma$ spectra to the detailed subthreshold behavior of the $\bar{K}N$ interaction. The $\Lambda_{c}$ weak decay is shown to enhance the signal of the $\Lambda(1405)$ because of the isospin filter mechanism that is characteristic of this decay.

\item[3.] The $\bar{K}N$ molecular picture of the higher energy pole of the $\Lambda(1405)$ is verified in two ways. The realistic $\bar{K}N$ potential generates a spatially extended structure of the $\Lambda(1405)$, in accordance with the loosely bound two-body molecular picture. The $\bar{K}N$ compositeness of the $\Lambda(1405)$ is evaluated by the model-independent relation and the threshold observables found in chiral SU(3) dynamics, indicating the dominance of the $\bar{K}N$ composite component in $\Lambda(1405)$.

\end{enumerate}
Accurate constraints provided by precision measurements exploring the dynamics of the coupled $\bar{K}N$ and $\pi\Sigma$ channels have been vital in order to reach these conclusions. Active cooperation of theory and experiments is expected to promote this area of reseach further in the near future.

\section*{Acknowledgements} 

This work was partly supported by the Grants-in-Aid for Scientific Research on Innovative Areas from MEXT (Grant No. 2404:24105008),
by RIKEN Junior Research Associate Program, 
by RIKEN iTHES Project,
by the Yukawa International Program for Quark-Hadron Sciences (YIPQS),
by Open Partnership Joint Projects of JSPS Bilateral Joint Research Projects, 
by JSPS KAKENHI Grants Nos. 23224006, 24740152 and 25800170,
by DFG through CRC 110,
by the Spanish Ministerio de Economia y Competitividad, European FEDER funds under Contract No. FIS2014-51948-C2-1-P and FIS2014-51948-C2-2-P, 
by the Generalitat Valenciana in the program Prometeo II, 2014/068,
and by the European Community-Research Infrastructure Integrating Activity Study of Strongly Interacting Matter (acronym HadronPhysics3, Grant Agreement n. 283286) under the Seventh Framework Programme of the EU. 








\end{document}